# A Superconductor Made by a Metal Heterostructure at the Atomic Limit Tuned at the "Shape Resonance": MgB$_2$

A. Bianconi*, D. Di Castro, S. Agrestini, G. Campi, N. L. Saini,
*Unità INFM and Dipartimento di Fisica, Università di Roma "La Sapienza",
P. le Aldo Moro 2, 00185 Roma, Italy*

A. Saccone, S. De Negri, M. Giovannini
*Dipartimento di Chimica e Chimica Industriale, Università di Genova,
Via Dodecaneso 31, 16146 Genova, Italy*

We have studied the variation of superconducting critical temperature $T_c$ as a function of charge density and lattice parameters in $Mg_{1-x}Al_xB_2$ superconducting samples. The $AB_2$ heterostucture of metallic boron layers (intercalated by A=magnesium, aluminum layers, playing the role of spacers) is made by direct chemical reaction. The spacing between boron layers and their charge density are controlled by chemical substitution of Mg by Al atoms. We show that high $T_c$ superconductivity is realized by tuning the chemical potential at a "shape resonance" according with the patent for "high-temperature superconductors made by metal heterostructures at the atomic limit". The energy width of the superconducting shape resonance is found to be about 400 meV.

*For further information:* http://www.superstripes.com/

*Corresponding author, email: antonio.bianconi@roma1.infn.it

The material design of heterostructures is reaching the atomic limit for engineering the electronic, transport [1] and optical [2] properties of materials. The electronic energy levels, wave functions and band structure of new materials can now be designed by quantum engineering. The basis of quantum engineering is the control of the quantum confinement of the electron gas in mesoscopic units to obtain artificial discrete electronic states in potential quantum wells (i.e., the classical quantum effects for a particle in a box). While most of the interest has been focused on semiconductor heterostructures, now the interest is shifting to metallic heterostructures in order to design and tailoring in unprecedented ways the superconducting properties [3]. The size of the mesoscopic units in these heterostructures is of the scale of the Fermi wavelength $\lambda_F=2\pi/k_F$ (where $k_F$ is the wavevector of electrons at the Fermi level). While in a typical semiconductor heterostructure, as gallium arsenide at room temperature, the Fermi wavelength $\lambda_F$ is of the order of ~25 nm, in a metallic heterostructure $\lambda_F$ reaches the atomic limit in the range ~0.3-1.5 nm. The atomic monolayers provide the thinnest possible metallic slabs (also called quantum wells) where the electronic charge is delocalized over an effective thickness $L^*$ of the order of the Fermi wavelength [4]. A superconducting heterostructure [5] is made by a superlattice of quantum wells of a first material B intercalated between a second material, A, that provides a periodic potential barrier for the electrons in the first material. The units BA are repeated a large number of times to form a superlattice ... BABABA ... with period $\lambda_p$. A superlattice of atomic monolayers can be realized by planes (with a bcc or hcp, or fcc, or, honeycomb crystallographic lattice) of atoms B alternated with other planes (with a different bcc or hcp, or fcc, or, honeycomb crystallographic lattice) of atoms A rotated one respect the others to get a lattice matching [6]. Stable pseudomorphic lattices of A or B materials appear in the heterostructures that are not stable in the separated homogenous materials. The internal stress due to lattice mismatch between the A and B intercalated layers induces a local micro-strain on each layer. The anharmonicity of lattice dynamics and multi-phonons interaction in the metallic layers are controlled by the micro-strain.

The growth of artificial metallic quantum superlattices at the atomic limit has been delayed by the technical problems for epitaxial growth of thin homogenous films with sharp interfaces at the atomic limit. In fact, the metallic thin films, grown by standard evaporation methods, get <u>disordered</u> on an atomic scale or <u>granular</u> on a mesoscopic scale and electrons get <u>localized</u> at low temperature.

Atomically perfect superlattices can be grown by direct chemical synthesis of elemental atoms A and B in optimized thermodynamic conditions. The superlattices of quantum wells at the atomic limit shown in Fig. 1, were known since 1935 as the $AlB_2$ crystallographic structure [7-10]. Let us consider the case of A=aluminum or magnesium and B=boron; these two elements do not mix in the solid phase and the radius of B is much smaller than the radius of A=Mg (Be or Al). However at high temperature (T=1223K) Mg atomic gas diffuses inside the boron crystals and a superlattice of pseudomorphic honeycomb graphite-like boron monolayers with intercalated hexagonal, Mg (Al, or Be), monolayers is formed.

The use of these type of metallic superlattices as superconductors at high temperature to overcome the technical limitation of cuprate perovskites and standard superconducting Nb alloys has been described in a patent [11] that discloses how the superconducting temperature can be amplified by a large factor. The amplification is achieved in a superlattice of quantum metallic wells by tuning the chemical potential (the Fermi energy $E_F$) at a "shape resonance" occurring near the quantum critical point (QCP) [12-22] at energy $E_c$ for a 2D-3D transition near the threshold (bottom or top) of the 2D subbands of the superlattice. The superconducting critical temperature ($T_{cn}$) is amplified by an amplification factor f of the order of 100-1000 from the low temperature range to the high temperature range by tuning the Fermi energy over an energy range $|E_c - E_F| < \Delta_0$, with $\Delta_0 > t_z$, where $t_z$ is the small electron hopping energy between the boron layers.

It was known that in the $AlB_2$-type diborides the electronic structure of the boron superlattice is formed by two dimensional (2D) boron $\sigma(2p_{x,y})$ subbands that coexist with other 3D $\pi$ $(2p_z)$ bands crossing the Fermi level [23]. Near the top of the 2D boron $\sigma(2p_{x,y})$ subband, as shown Fig. 2, the characteristic QCP of the superlattice occurs at the energy $E_c$ at the $\Gamma$ point where a transition to the 3D regime occurs for $E > E_c$ up the top of the subband at A point. The electrons in the B $\sigma(2p_{x,y})$ subband have cylindrical Fermi surface in the energy range $E_F < E_c$ where hopping between the metallic units in the transversal direction is not allowed and a closed 3D Fermi surface in the energy range at $E_c < E_F < E_c + 4t_z$ (see Fig. 2) where hopping between the metallic layers is allowed giving a the small dispersion perpendicular to the metallic layers with bandwidth $4t_z$.

According with ref. [11] the high $T_c$ superconductors can be made by diborides if the chemical potential is tuned for $T_c$ amplification in the energy range $|E_c - E_F| < \Delta_0$. The "shape resonance" can be obtained by tuning the energy level $E_F$ (via changes of the charge density) and/or tuning the energy position of $E_c$ (via changes of the lattice parameters of the superlattice). The difficulty to reach the resonance point is due to the fact that the resonance is very narrow. Therefore both the lattice parameters and the charge density of metallic monolayers have to be controlled in a very fine scale ($\Delta_0 << E_F$ in good metals) for a fine tuning of the energy difference $\Delta = E_F - E_c$.

In most of diborides the Fermi level is out of resonance $|E_F - E_c| > \Delta_0$ but in $MgB_2$ the Fermi level is close to $E_c$ [23] therefore $T_c$ is amplified according with the claims in ref. 11. $MgB_2$ was used as a component for chemical reactions but never tested for superconducting properties before january 2001 when it has been shown to be a high $T_c$ superconductor [24-26] with promising applications for energy transport and in electronics according with ref. 11. $T_c$ reaches 40 K with an amplification of a factor f~100 in $MgB_2$ in comparison with other diborides where $T_c$<0.4K. The shape resonance occurs for critical values of the lattice parameters (a and c that control the position of $E_c$) and of the charge density (that controls $E_F$ ) [27].

Here we report an investigation of the high $T_c$ superconducting phase in $MgB_2$ by changing both the charge density and the superlattice structural parameters. In order to tune the chemical potential $E_F$ around the quantum critical point at $E_c$ in the energy range $|E_F - E_c| < 400 meV$ we have used the chemical substitution of Mg by Al atoms in the $Mg_{1-x}Al_xB_2$ samples. We show that high $T_c$ superconductivity is realized at a "shape resonance" by tuning the chemical potential within an energy range $\Delta_0$ = 400 meV from the QCP.

The powder $Mg_{1-x}Al_xB_2$ samples have been synthesized by direct reaction of the elements. The starting materials were elemental magnesium and aluminum (rod, 99.9 mass% nominal purity) and boron (99.5 % pure <60 mesh powder). The elements in a stoichiometric ratio were enclosed in tantalum crucibles sealed by arc welding under argon atmosphere. The Ta crucibles were then sealed in heavy iron cylinder and heated for one hour at 800 °C and two hours a 950 °C in a furnace. The samples were characterized by x-ray diffraction and the lattice parameters were determined by standard least-squares refinement of the diffraction data recorded using the Cu $K_\alpha$ emission at room temperature. The Al doping in $Mg_{1-x}Al_xB_2$ samples controls: 1) the spacing between the metallic boron layers (measured by the c-axis); 2) the in plane B-B distance (measured by the a-axis); and 3) the charge density that increases by x electrons per unit cell. The variation of the ratio $a(x)/a_0$ and $c(x)/c_0$ where $c_0$ and $a_0$ are the crystallographic axis in undoped $MgB_2$ compound as a function of aluminum content x are plotted in Fig. 4. By increasing Al content x we observe a contraction of the spacing between the boron monolayers shown in Fig. 1 given by the variation of the c-axis. Fig. 4 shows that the variation of the **a**-axis with aluminum doping is much smaller than that of the **c** axis as observed by pressure effect [29]. The contraction of the a-axis measures the reduction of the micro-strain of the B-B distance with Al substitution [27]. The expected variation of the lattice parameters for a solid solution of $AlB_2$ and $MgB_2$ are indicated by the dashed lines. We observe a variation of the slope of both lattice parameters at aluminum content x=0.3.

The superconducting properties of $Mg_{1-x}Al_xB_2$ superconductors have been investigated by the temperature dependence of the complex conductivity by the single-coil inductance method [30]. This method is based on the influence of the sample on the radio frequency complex impedance of a LC circuit. Temperature-dependent measurements of the complex impedance, with and without the sample, allow extracting the complex conductivity of the sample. The real part of the complex conductivity is related with the dissipative conductivity, while the imaginary part is a measure of the London penetration depth. Therefore, the temperature dependence of the imaginary part contains information about the superfluid density and dissipation effects in the vortex dynamics.

The sample was located in the vicinity of the coil. A sketch of the experimental circuit geometry is shown as an inset of Fig. 5 that shows the ratio of the square of resonant frequencies $\frac{f_0(T)^2}{f(T)^2}$ where $f_0(T)$, and $f(T)$, are the LC resonance frequency measured without and with the sample, respectively. Below the superconducting transition, the inductance L of the circuit decreases because of the screening sheet currents, and, hence, the resonant frequency f(T) shows a sharp increase. Therefore at the transition temperature $T_c$ a sharp drop of the ratio $\frac{f_0(T)^2}{f(T)^2}$ is observed that is a good probe of the superconducting transition. The derivative of $\frac{f_0(T)^2}{f(T)^2}$ is shown in the upper panel of Fig. 5 and its derivative maximum has been used to define $T_c$. The results show that we have succeeded to change by aluminum doping two key parameters that control the critical temperature: 1) The charge density that has been increased up to 0.5 electrons/cell rising up the chemical potential $E_F(x)$; 2) The lattice parameters **c** and **a** that control the energy shift of $E_c(x)$ in the electronic structure of the superlattice of boron monolayers,

According with band structure calculations for $Mg_{1-x}Al_xB_2$ at x=0 in the pure system the Fermi level $E_F$ is in the 2D regime at 395 meV below the QCP of the σ $2p_{x,y}$ band at $E_c$. The 3D regime occurs over an energy range $4t_z$=368 meV above $E_c$. The decrease of the a-axis, a(x), shown in Fig. 4 pushes to higher energy the position of the critical point $E_c(x)$ at Γ. The decrease of the **c**-axis with Al doping shown in Fig, 4 increases the dispersion perpendicular to the metallic boron monolayers $t_z$ and decreases the position of $E_c(x)$. The position of the Fermi level $E_F(x)$ changes with x due to changes both of the electron counts and of the density of states. At aluminum doing x=0.5 the σ $2p_{x,y}$ is nearly filled and the Fermi level is in the 3D regime above the QCP at $\Delta(0.5) = E_c(0.5) - E_F(0.5) = -320 meV$ [23]. Therefore by aluminum doping in the range 0<x<0.5 range it is possible to test the variation of the critical temperature at the shape resonance by changing the chemical potential in the range 320<Δ(x)<395 meV.

In Fig. 6 we report the variation of the critical temperature as a function of the energy shift of the chemical potential with Al doping x. The energy shift of the chemical potential Δ(x) as a function of the aluminum doping x has been obtained by Massidda et al. from their band structure calculations [23].

The critical temperature follows qualitatively the density of states as expected for a shape resonance [18,19]. The high $T_c$ occurs by tuning the Fermi level over a range $\Delta_0$=400 meV above and below the QCP at $E_c$. High critical temperature 23K<$T_c$<40K occurs for $E_F$<$E_c$ in the 2D regime and the low critical temperature 5K<$T_c$<13K occurs in the 3D range for $E_c$<$E_F$.

In conclusion we have shown that the $MgB_2$ superconductor with high critical temperature in accordance with reference 11, is characterized by being formed by a plurality of first portions formed of first layers (of graphite-like structure) made by a first superconducting material (boron), alternated with second portions formed by second layers (of hexagonal structure) made of a second material (magnesium) with different electronic structure; said first portions being formed of first layers having a thickness L, and said second portions being formed by second layers having a thickness W, with a period L+W=c such as to tune the Fermi level to a "shape resonance" of the superlattice. The high critical temperature is achieved by tuning the chemical potential relative to the QCP at $E_c$ for 2D to 3D transition in the $\sigma$ $2p_{x,y}$ by changing both the lattice parameters and the charge density in $Mg_{1-x}Al_xB_2$.

The width of the shape resonance has been found to be of the order of 400 meV larger than the transverse band dispersion $4t_z$ and several time the energy of the $E_{2g}$ phonon in the range of 600-800 $cm^{-1}$. This can be associated with strong anharmonic effects due to the large micro-strain of the B-B distance in the metallic boron layers [27] as in cuprates [31]. These results implies a pairing mechanism mediated by many phonons and coupled with electronic excitations where the chemical potential is tuned at a shape resonance [11]. The high $T_c$ in $MgB_2$ shows up in a narrow regime near critical values of the lattice parameters of the superlattice at the atomic limit and at a critical density according with [11] therefore other new heterostructures at the atomic limit of similar type could be synthesized by chemical reactions to get higher $T_c$. via the shape resonance effect with the possibility to reach room temperature superconductors.

We would like to thank S. Massidda, G. Profeta, A. Perali, and J. B. Neaton for useful discussions and information on the shift in the electronic band structure with aluminum doping. This research has been supported by i) Progetto 5% Superconduttività del Consiglio Nazionale delle Ricerche (CNR); ii) Istituto Nazionale di Fisica della Materia (INFM); and iii) the "Ministero dell'Università e della Ricerca Scientifica (MURST).

REFERENCES

1. S. Datta *Electronic Transport in Mesoscopic Systems* (Cambridge University, Cambridge, 1996).

2. F. Capasso, J. Faist, and C. Sirtori *J. Math. Phys.* **37**, 4775 (1996).

3. M. Brooks, *New Scientist* No.2192, 38, 26 Jun1999; A. *Bianconi Physics World* 11, N0.7 pag. 20 (July 1998)

4. I. P. Batra S. Ciraci, G. P. Scrivastava J.S. Nelson, and C.Y. Fong *Phys. Rev. B* **34**, 8246 (1986); S. Ciraci and I. P. Batra *Phys. Rev. B* **33**, 4294 (1986).

5. S. T. Ruggiero and M. P. Beasley in *Synthetic Modulated Structures* edited by L. L. Chang and B.C. Giessen, New York Academic Press (1985) pag. 365.

6. F. J. Himpsel, *Phys. Rev B* **44**, 5966 (1991); J. E. Ortega, F. J. Himpsel G. L. Mankey, and R. F. Willis *Phys. Rev. B* **47**, 1540 (1993); G. L. Mankey, R. F. Willis and F.J. Himpsel, *Phys. Rev. B* **47**, 190 (1993); P. Segovia, E. G. Michel and J.E. Ortega, *Phys. Rev. Lett.* **77**, 3455 (1996).

7. W. Hoffmann and W. Jänicke *Naturwiss.* **23**, 851 (1935); *Z. Physik. Chem.* **31** B, 214 (1936).

8. V. Russel, R. Hirst, F. Kanda, and A. King *Acta Cryst.* **6**. 870 (1953).

9. E. Jones and B. Marsh *J. Am. Chem. Soc*. **76**, 1434 (1954).

10. P. Duhart *Ann. Chim. t.***7***,* 339 (1962).

11. "*High $T_C$ superconductors made by metal heterostuctures at the atomic limit*; (priority date 7 Dec 1993), International patent PCT WO 95/16281 (15 June 1995); European Patent N. 0733271 published in the *European Patent Bulletin* 98/22, (1998}; Japanese patent JP 286 8621; and web site http://www.superstripes.com/.

12. A. Bianconi *Sol. State Commun.* **89**, 933 (1994).

13. A. Bianconi, M. Missori, N. L. Saini, H. Oyanagi, H. Yamaguchi, D. H. Ha, and Y. Nishiara *Journal of Superconductivity* **8**, 545 (1995).

14. A. Perali, A. Bianconi, A. Lanzara, and N. L. Saini *Solid State Communications* **100**, 181 (1996).

15. A. Perali, A. Valletta, G. Bardelloni, A. Bianconi, A. Lanzara, N.L. Saini *J. Superconductivity* **10**, 355-359 (1997) (Proc. of the First Inter. Conference on *Stripes, Lattice instabilities and high $T_c$ Superconductivity* Roma Dec 1996).

16. A. Valletta, G. Bardelloni, M. Brunelli, A. Lanzara, A. Bianconi, N.L. Saini *J. Superconductivity* **10**, 383-387 (1997), (Proc. of the First Inter. Conference on *Stripes and high $T_c$ Superconductivity* Roma Dec 1996).

17. A. Bianconi, A. Valletta, A. Perali, and N.L. Saini *Solid State Commun.* **102**, 369 (1997).

18. A. Valletta, A. Bianconi, A. Perali, N. L. Saini *Zeitschrift fur Physik B* **104**, 707 (1997).

19. A. Bianconi, A. Valletta, A. Perali, and N. L. Saini *Physica C* **296**, 269 (1998).

20. A. Bianconi, S. Agrestini, G. Bianconi, D. Di Castro, N. L. Saini, in Stripes and Related Phenomena, A. Bianconi N. L. Saini Editors, Kluwer Academic/Plenum publisher, New York, 2000 p. 9-25 (Proc. of the second Int. Conference on *Stripes and high $T_c$ Superconductivity* Rome Italy, Jun 1998).

21. A. Bianconi *Int. J. Mod. Phys. B,* **14**, 3289 (2000) (Proc. of the third Inter. Conference on *Stripes and high $T_c$ Superconductivity* Rome, Italy Sept 2000).

22. N. L. Saini and A. Bianconi *Int. J. Mod. Phys. B,* **14** 3649 (2000) (Proc. of the third Int. Conference on *Stripes and high $T_c$ Superconductivity* Rome Italy Sept 2000).

23. G. Satta, G. Profeta, F. Bernardini, A. Continenza, S. Massidda; cond-mat/0102358 (2001); references therein and private communication.

24. J. Nagamatsu, N. Nakagawa, T. Muranaka, Y. Zenitani, and J. Akimitsu, Nature **410**, 63 (2001)

25. S. L. Bud'ko, S. L. Lapertot, C. Petrovic, C. E. Cunninghamy, N. Anderson, and P. C. Canfield; Phys. Rev. Lett. **86**, 1877 (2001).

26. D. E. Finnemore, J. E. Ostenson, S. L. Bud'ko, G. Lapertot, and P. C. Phys. Rev. Lett. **86**, 2420 (2001).

27. A. Bianconi, N. L. Saini D. Di Castro, S. Agrestini, G. Campi, A. Saccone, S. De Negri, M. Giovannini, M. Colapietro. cond-mat / 0102410 (2001).

28. J. S. Slusky, N. Rogado, K. W. Reagan, M. A. Hayward, P. Khalifah, T. He, K. Inumaru, S. Loureiro, M. K. Hass, H. W. Zandbergen, R. J. Cava Nature **410**, 343 (2001).

29. J. D. Jorgensen, D. G. Hinks, and S. Shortl cond-mat/ 0103069 (2001).

30. D. Di Castro, N. L. Saini, A. Bianconi, A. Lanzara *Physica C* **332**, 405 (2000).

31. A. Bianconi, G. Bianconi, S. Caprara, D. Di Castro, H Oyanagi, N. L. Saini, *J. Phys.: Condens. Matter,* **12** 10655 (2000).

**FIGURE CAPTIONS**

**Fig. 1.**   Pictorial view of the superlattice of metallic boron monolayers (B) made of graphite-like honeycomb lattice separated by hexagonal Mg layers (A) forming a superlattice ABAB in the c-axis.

**Fig. 2.**   The superconducting nth shape resonance occurs near quantum critical point at energy $E_c$ at the top of a single particular hole like subband of the electronic structure of a superlattice of quantum wells where the partial electronic structure of this subband shows a dimensional transition from 2D at $E<E_c$ to 3D in the range $E_c<E<E_0$.

**Fig. 3.**   Pictorial view of the transition in a particular subband of a superlattice of quantum wells, from a 2D-like Fermi surface below $E_c$ to a 3D-like Fermi above $E_c$. The shape resonance for the superconducting gap occurs in a energy range $\Delta_o$ around $E_c$.

**Fig. 4.**   The variation of the ratio a and c axis: $a(x)/a_0$ and $c(x)/c_0$ as a function of Al doping x (where =$a_0$ and $c_0$ are the lattice parameters of the $MgB_2$ compound) in $Mg_{1-x}Al_xB_2$ samples.

**Fig. 5.**   *Lower panel*: the radio frequency surface resistance probed by the ratio $\frac{f_0(T)^2}{f(T)^2}$ of the probing LC circuit of $Mg_{1-x}Al_xB_2$ samples. The inset shows a cartoon sketch of the experimental system. *Upper panel*: the peak of the derivative of the ratio $\frac{f_0(T)^2}{f(T)^2}$ at $T_c$.

**Fig. 6.**   The critical temperature $T_c$ as a function of the shift of the chemical potential $\Delta(x)= E_c(x)-E_F(x)$, where $E_F$ is the Fermi level and $E_c$ the QCP of the boron $\sigma$ $2p_{x,y}$ band, where a 2D to 3D transition occurs, as a function of Al doping x.

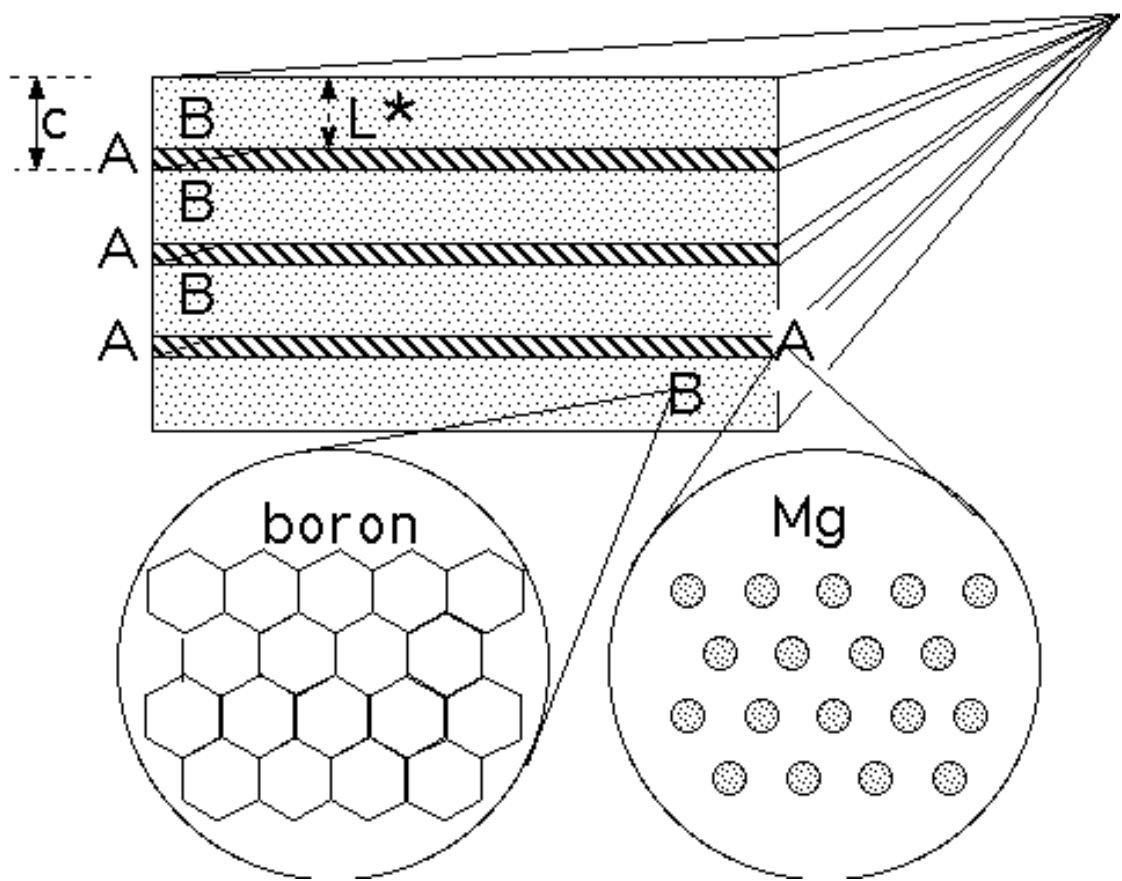

Fig. 1

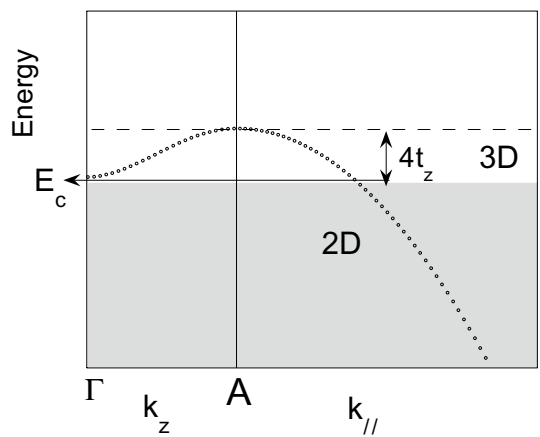

Fig. 2

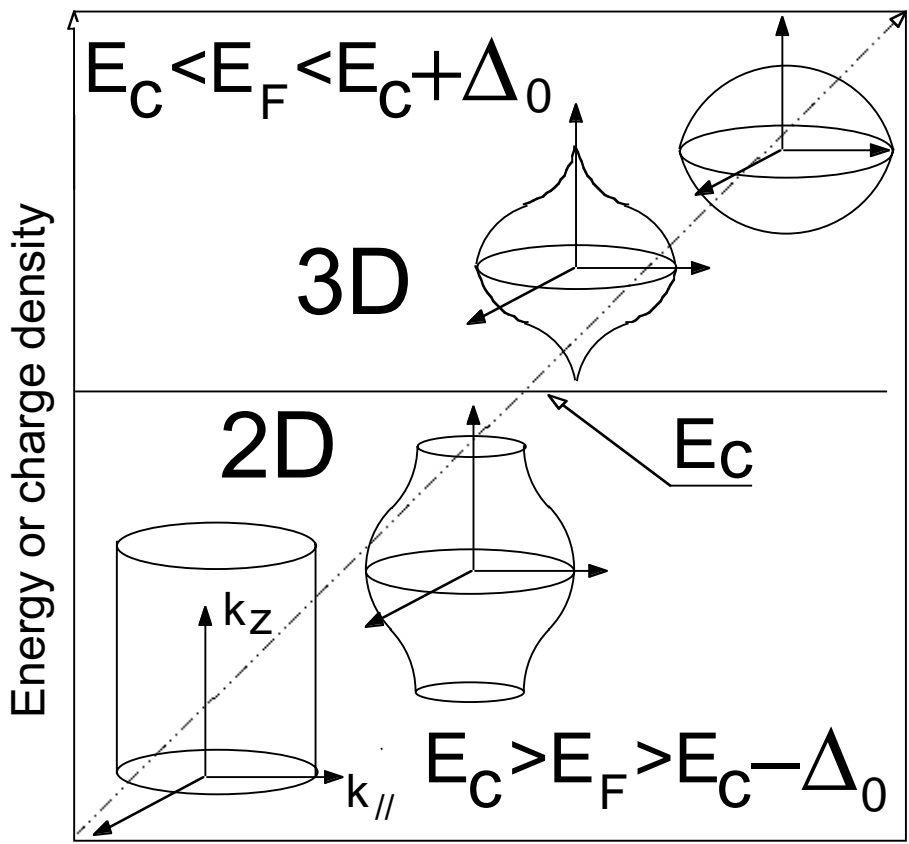

Fig. 3

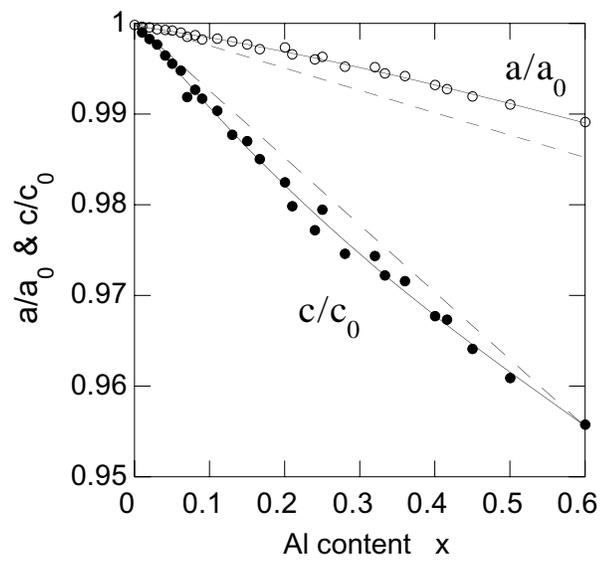

Fig. 4

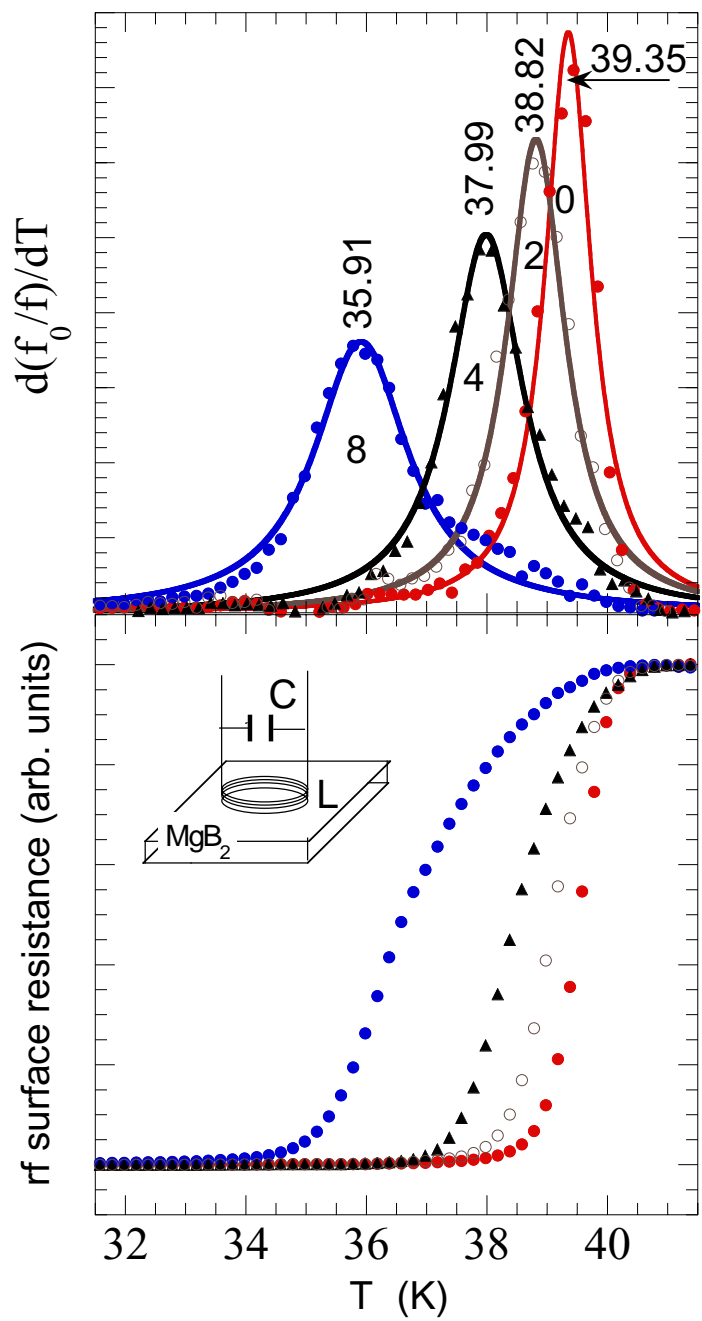

Fig. 5

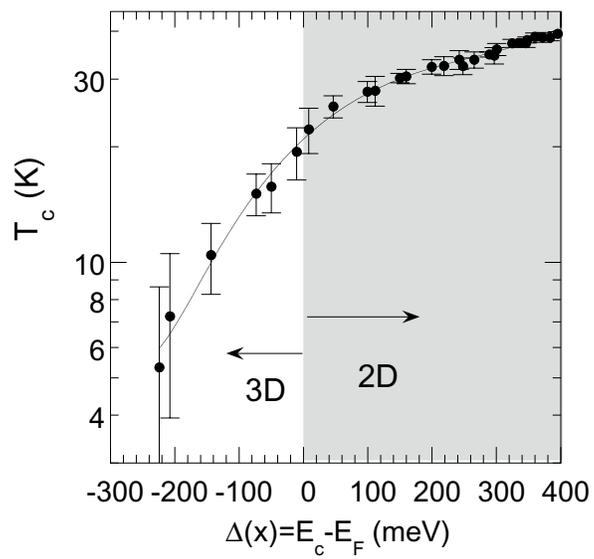

Fig. 6